# Metalens With Artificial Focus Pattern

## Authors
Mao Ye[1], Vishva Ray[2], Dachuan Wu[1], Yasha Yi[1,3]*

## Affiliations
[1] Integrated Nano Optoelectronics Laboratory, University of Michigan, 4901 Evergreen Rd., Dearborn, Michigan, 48128, USA

[2] Lurie Nanofabrication Facility, Department of Electrical Engineering and Computer Science, University of Michigan, Ann Arbor, Michigan 48109-2122, USA.

[3] Energy Institute, University of Michigan, 2301 Bonisteel Blvd., Ann Arbor, Michigan, 48109, USA.

*E-mail: yashayi@umich.edu

## Abstract

**Metalens as one of the most popular applications of emerging optical metasurfaces has raised widespread interest recently. With nano structures fully controlling phase, polarization and transmission, metalens has achieved comparable performance of commercial objective lenses. While recent studies seeking for the accomplishment of traditional focusing behaviors through metalens are successful, in this work, we have discovered that instead of focusing light to a point, metasurface further enables shaping the focus into a flexibly designed pattern, with more promises and potentials. New mechanism and generalizations of conventional point-focused metalens guiding principles have been proposed with metalens concentrating light to artificial focus pattern. As proving examples, we have demonstrated the engineering of metalens with artificial focus pattern by creating line and ring-shaped focus as 'drawing tools'. The metalens with 'U' and 'M' shaped focus are characterized for the proof of concepts. These metalens are fabricated through a single layer of silicon-based material through CMOS compatible nano fabrication process. The mechanism to generate artificial focus pattern can be applied to a plethora of future on-chip optical devices with applications ranging from beam engineering to next generation nano lithography.**

## Introduction

Emerging optical metasurfaces comprised of a thin layer of nano structures are reported with exceptional control over wavefront [1-3]. Through precise engineering of phase, polarization and amplitude, metasurfaces enabled miniaturization of traditional bulk optical devices to thin planar structures. The transformation of traditional focusing lens into a metalens [4-10] has attracted intense interest due to its widespread application in the field of imaging and sensing (cameras, microscopy, displays and sensors). To date, metalens have been reported with comparable performance of a commercial objective including functions of achromatic [11, 12], wavelength tunability [13], and polarization insensitive [14] /distinguishing [15] characteristics. Basically, previous studies are successful on achieving functionalities of traditional lens or lens sets through careful engineering of nano phase shifters. While in this work, we have discovered that these nano structures (e.g. metalens) have far more promises and potentials than just accomplishing traditional 'point' focusing behavior.

Here, counterintuitively, we have created metalens that enables focusing light onto an 'artificial and flexible designed pattern' instead of a point. The shaping of focus is achieved by introducing extra spatial information into the phase reconstruction formula that enabled focusing light into a pattern with designed focal length. As a result, the focus pattern created shares the property of a focus such as diffraction limitation. This is a completely novel and different approach, compared with other metasurface based pattern-generation method such as holography [16, 17], where pattern is created by recording (iteration) of an existing pattern's phase profile. In order to successfully achieve the artificial focus patterns and demonstrate our proof of concepts, we have utilized both grating- and cylinder-based phase shifter design to show their feasibilities in applications on different targeting patterns. Two metalens are designed and fabricated with focus pattern 'U' and 'M' respectively for the proof of concept. The average numerical aperture (NA) is 0.8 (NA may vary at different locations on the focus pattern), and the size is 40×60 µm for 'U'-shaped focus metalens and 44×80 µm for 'M'-shaped focus metalens. Both lenses are characterized under 685 nm incident light and the designed patterns are generated 15 µm away from lens (focal length = 15 µm).

Metalens are designed through space discretization and phase construction process [14]. Basically, the lens plane is first discretized into subwavelength sized pitches (period). Within each period, high refractive index material is filled with varied fill factor to tune the effective index (or Pancharatnam-Berry phase) which ultimately controlled the phase shift of this region (e.g. generation of dimension variant phase profile). Conventional point-focused metalens can be designed by matching dimension-variant phase profile to the space-variant phase profile with focusing effect (Eq. 1) [4].

$$\phi(x,y) - \phi(0,0) = \frac{2\pi}{\lambda}\left(\sqrt{x^2 + y^2 + f^2} - f\right) \quad (1)$$

Where $x$, $y$, $\lambda$, and $f$ are space coordinate, incident wavelength and focal length. In this way we focus light as a point. While recent studies seeking for the accomplishment of traditional focusing behaviors through metalens, we discovered that instead of focusing light to a point, metasurface enabled focusing light into a designed pattern. The method of artificial focus pattern brings extra flexibility to optical metasurface can potentially be applied to a plethora of future applications.

## Results

**Principle of artificial focus pattern design.** In order to focus light into a pattern, extra spacial relationship is induced into the phase profile. For a line-shaped focus with controlled orientation, the phase profile can be represented by Eq. (2) and Eq. (3).

$$\phi(x,y) - \phi(0,0) = \frac{2\pi}{\lambda}\left(\sqrt{(x + a(y))^2 + f^2} - f\right) \quad (2)$$

$$-a(y) = \frac{y-c}{k} \quad (3)$$

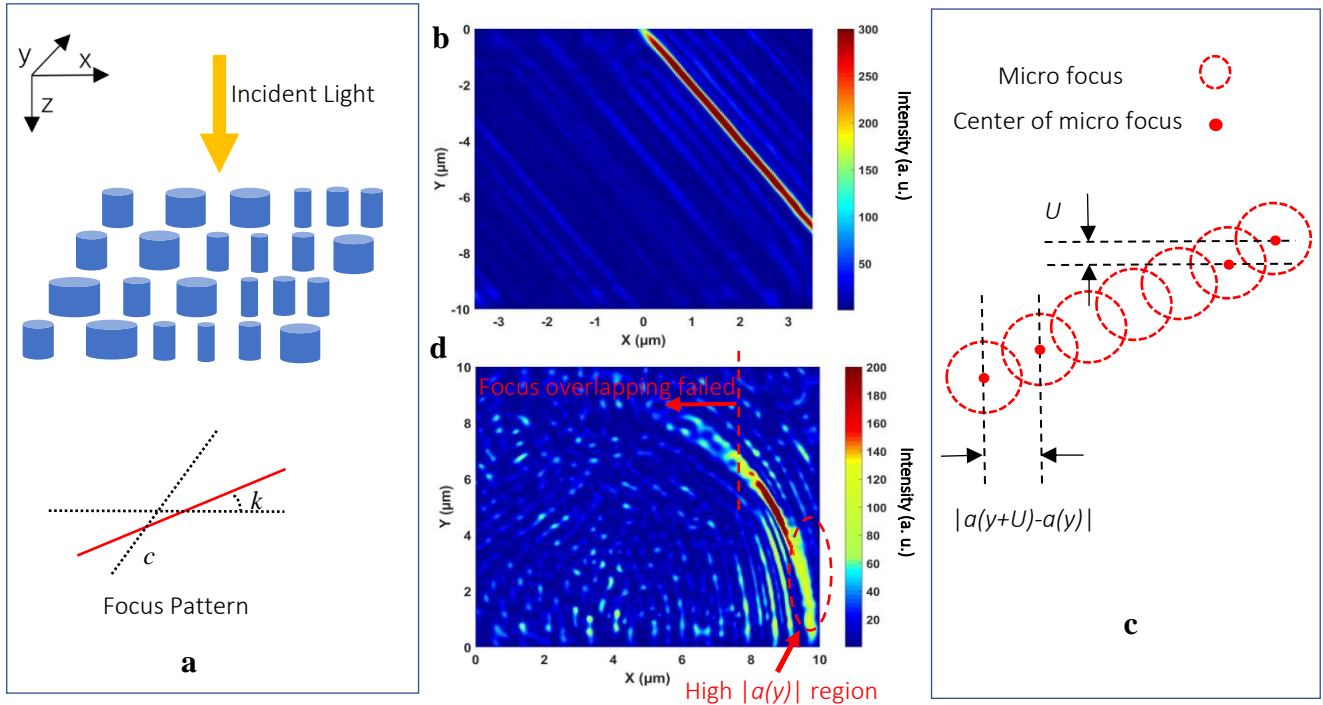

Fig. 1 Design and discussion of line-shaped focus pattern. (a) Schematic graph of metalens creating line-shaped focus. (b) FDTD simulation of typical example of line-shaped focus by combining Eq. (2) and Eq. (3) with $k=-2$, $c=0$. The detailed structure-phase relationship is provided at supplementary figure (S1). (c) Illustration of the mechanism of creating focus pattern through 'micro focus' overlapping. (d) FDTD simulation of arc shaped focus by combining Eq. (2) and Eq. (5).

Where *k* and *c* are the slope and spacial constant of line-shaped focus at *x-y* focal plane (Fig.1 a), *a(y)* is a function of the coordinate *y*. Basically, Eq (2) is creating focus with off-axis parameter *a(y)* and this parameter is a function of coordinate *y*. An example of line-shaped focus created by combining Eq. 2 and Eq. 3 is shown in Fig. 1 (b) where a line-shaped focus is created with *k*=-2, *c*=0. The metalens (10×10 µm sized) is designed through propagation phase with cylindrical shaped nano structure (Fig. 1 a), and the pattern is created 6 µm away from lens plane (detailed phase-structure information is shown in S1).

We can intuitively understand this focus pattern (Fig. 1 b) as a result of 'overlapping' of multiple 'micro focuses'. To be specific, based on Eq. (3), the function *a(y)* is varied based on the the phase shifter's *y* coordinate. In this case, phase shifters with same *y* coordinate (e.g. each line) can be considered as a 'micro metalens'. These metalens generate off-axis focus based on the value of *a(y)*. Due to close distance between adjacent focuses, these focuses become overlapped and form continuous pattern, which is shown in Fig. 1 (c). The distance between adjacent 'micro focus' can be calculated as:

$$D(y) = \sqrt{(a(y+U) - a(y))^2 + U^2} \leq \frac{\lambda}{2NA(y)} \quad (4)$$

Where *U* (220 nm) and *NA(y)* are pitch (period) size and numerical aperture, *D(y)* is the distance between adjacent 'micro focus' created by adjacent 'micro metalenses'. Note that the *NA(y)* in Eq. (4) is the numerical aperture of the typical 'micro metalens', which is also a variable of *y* coordinate. And due to the diffraction limited size of 'micro focus', *D(y)* has to be smaller than the Abby's diffraction limit.

With this approach, we are able to create line-shaped focus across the focal plane (Fig 1 b). While, as we discovered, simply engineering *a(y)* (Eq. 3) will not render desired focus pattern even when the criterion of Eq. (4) is satisfied. To better illustrate this issue, we have designed an arc-shaped focus by modifying Eq. (3) to

$$-a(y) = \sqrt{r^2 - y^2} \quad (5)$$

Where *r* (10 µm) is the radius of the arc. This metalens is designed under same size (10×10 µm) and focal length (6 µm). The field distribution at focal plane is shown in Fig. 1 (d). In this example, the distance between adjacent focuses *D(y)* increased due to the increase of *y* coordinate, that the arc structure disappeared at *y* > 6 µm due to dissatisfaction of Eq. 5. However, counterintuitively, the intensity of arc does not follow an inverse relationship with *D(y)*. To be specific, the intensity of artificial focus pattern is closely related to *D(y)*, as smaller *D(y)* indicates closer packing of adjacent 'micro focuses' which intuitively may render higher intensity. While shown Fig. 1 (d), region with lower *D(y)* (*y* =0~3 µm, *D(y)*=2.4~68.4 nm) shows lower intensity compared with regions around *y*=5 µm (*D(y)*=132.9 nm).

To better understand the underline mechanism, we demonstrate part of metalens in 2D as Fig. 2 (a). Basically, metalens is designed through space discretization and phase reconstruction method, and the focusing performance is achieved through a discontinuous phase profile. By shrinking the pitch (period) size into deep subwavelength range, researchers are trying to minimize the influence induced through phase discontinuity. However, phase discontinuity always has an effect on the focusing performance, and this effect become more obvious for high NA metalens. As shown in Fig. 2 (a), ideal equiphase plane in 2D remains constant curvature (based on Eq. 1), and the ideal transmission angle $α_0(x)$ varies continuously as it is always perpendicular to equiphase plane. While for metalens, there exist step-phase variations over space (Fig. 2 a). If we assume equilibrium media within period (phase shifted within each period remains the same), then an average transmission angle ($α_n$) can be demonstrated by connecting the center of adjacent equiphase plane. As a result, the focus of metalens is formed by a number of finite-narrow beams with different transmission angle. And the width of each beam *dn* is given by

$$dn = \frac{U}{sinα_n} \qquad (6)$$

Where we assume the lens comprised of 2(N+1) periods and $α_n$ is the angle shifted between $n_{th}$ and $(n+1)_{th}$ period. Due to the limitation of material and nano fabrication, the pitch size *U* is cannot be infinitely small. As a result, for high NA metalens that comprises small $α_n$ region, *dn* can be a value much larger than the diffraction limitation. For example, if we consider a metalens designed under visible wavelength with *U*=0.35 µm (conventional size for $TiO_2$ based metalens), the largest beam width at its edge for NA=0.9 can render $d_N$=0.802 µm, and it reaches $d_N$=2.48 µm at NA=0.99. Large *dn* may render a much larger focus with decreased intensity depicted in Fig. 2 (b).

A metalens with off-axis focus can be considered as a combination of half-of-two metalenses with different NA but same focal length as shown in Fig. 2 (c). Due to the increasing value of *dn*, beam come from high NA part is concentrated to a larger area, which increased the size of focus and bring a decrease to focused power.

In our artificial focus pattern design, the focus is formed by a series of off-axis 'micro focus'. For focus generated with large *|a(y)|*, there exist a high NA part of metalens which rendered the effect depicted in Fig. 2 (c) (e.g. the focused power is reduced for 'micro metalens' comprised of large *|a(y)|* component). And this explained the issue we aforementioned for the arc-shaped focus (Fig. 1 d) where lower *D(y)* rendered lower focused power.

Basically, the diminishing of focus power problems are inherent drawbacks from space discretization, which can only be significantly improved when the pitch (period) size is reduced [18]. In addition, based on the mechanism discussed in Fig. 2, the focused intensity is varying as a function of *a(y)*. However, for the design of artificial focus pattern, it is important to maintain features with similar focused intensity across the whole pattern. In this application, we modified the Eq. (2) to

$$\phi(x,y) - \phi(0,0) = \frac{2\pi}{\lambda}\left(\sqrt{(x+a(y))^2 + (f+s(y))^2} - (f+s(y))\right) \qquad (8)$$

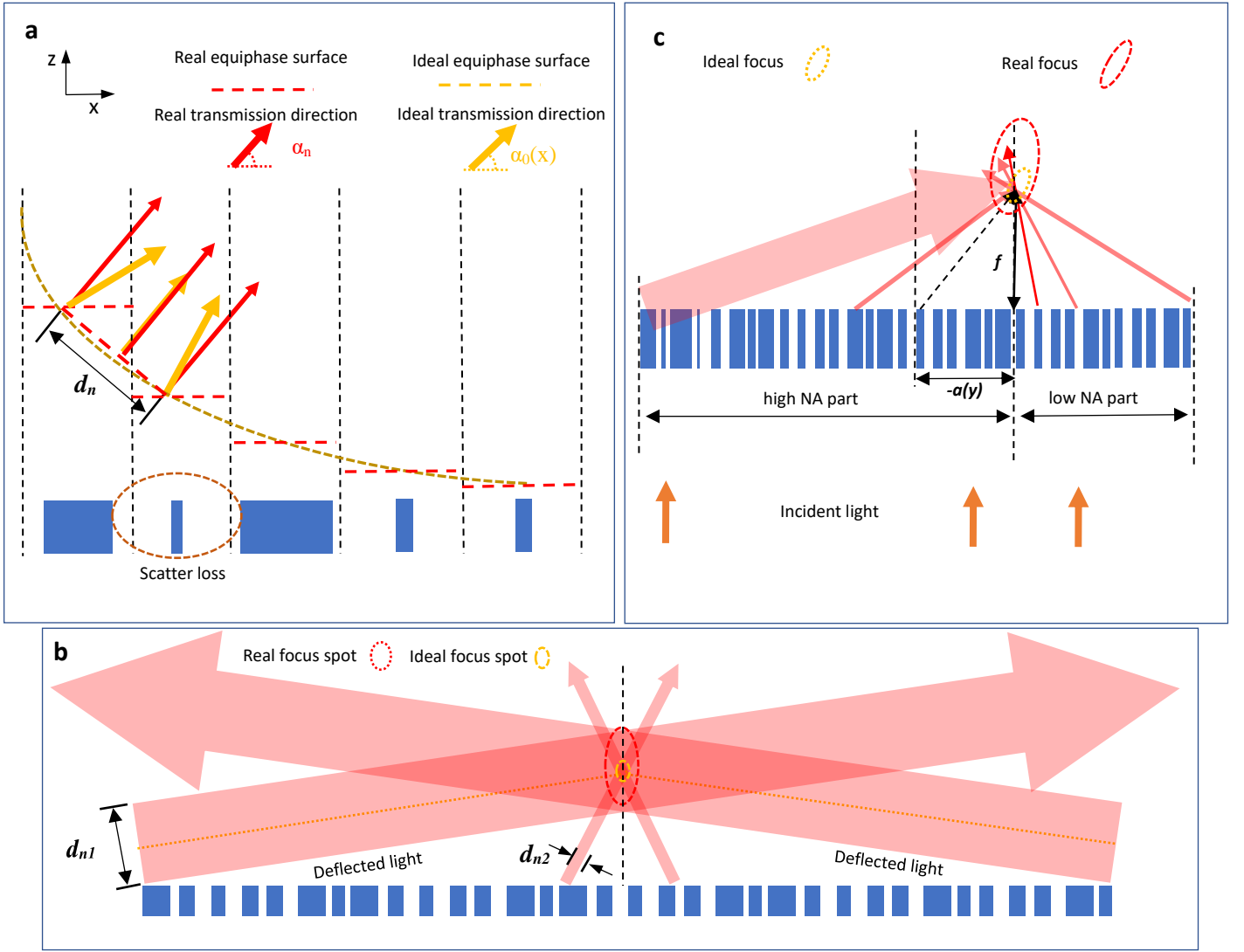

**Fig. 2** Focal shift mechanism of metalens. (**a**) Illustrative diagram of the metalens' focusing mechanism and the generation of phase discontinuity from space discretization. (**b**) Illustrative diagram of the influence of space discretization on focusing behaviors of metalens, where $d_{n1} \gg d_{n2}$ is the reason why real focus created by metalens is bigger than ideal focus. (**c**) Illustrative diagram of focusing mechanism for off-axis metalens, which is a basic element of our artificial focus pattern.

In Eq. (8), we add the focal shift component $s(y)$ as a modifier of focal length. The application of $s(y)$ is to tune the focused intensity for each 'micro metalens' in pursuit of a focused pattern with homogenous intensity at designed focal plane. The value of $s(y)$ can be acquired through simulations based on 'micro metalens' with different $a(y)$.

**Design and characterization of 'M' shaped focus.** A metalens with 'M'-shaped focus is designed using cylinder nano structure with feature size around 40 nm and thickness of 600 nm. The lens is fabricated with silicon rich silicon nitride material [18] and its structure is characterized through scan electron microscopy (SEM) and shown in Fig. 3 (a) and (b). The size of metalens is 44×80 μm with targeting 40×40 μm sized 'M' shaped focus.

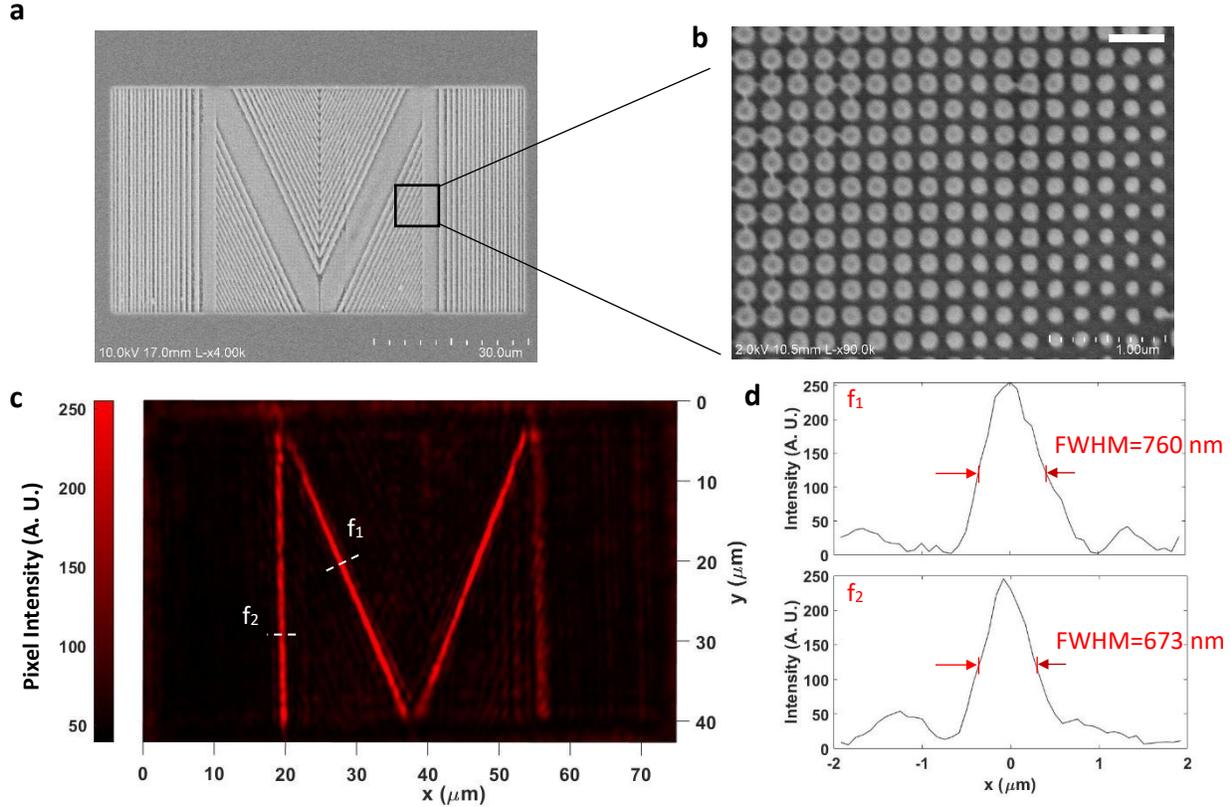

**Fig. 3** Structure and characterization of metalens with 'M' shaped focus. (**a**) SEM picture of the metalens's overall structure. (**b**) Magnified view of the lens which shows the structure of the cylinder phase shifter, size of scale bar is 500 nm. Both SEM picture are taken after gold sputter. (**c**) Characterized focal plane with incidence of 685 nm red light, two cross-sections are specified. (**d**) Focus profile at two cross-sections specified on 'M' shaped focus.

This lens is designed with 685 nm incident wavelength, and the whole pattern is aligned through Eq. (8). The characterized focus pattern is formed on the designed focal plane 15 μm away from lens (shown in Fig. 3 c). Basically, a clear 'M' shaped focus is formed at the designed focal plane (Fig. 3 c). One side line of 'M' is blur which may cause by defects of nano fabrication as can be observed in Fig. 3 (b) that some nano cylinder is interconnected.

Shown by Fig. 3 (c), the tilted line of 'M' exhibits a similar brightness, and this is caused by the adjustment of intensity from Eq. (8), that the intensity variation from varying $a(y)$ is aligned to a similar level. The focus profile at cross-sections in Fig. 3 (c) is shown at Fig. 3 (d). As shown, the full-width-at-half-maximum (FWHM) of both cross-sections are larger than the size of diffraction limited focus (NA varies from 0.6 to 0.8). This is due to the influence of space discretization as discussed in Fig. 2, that the size of off-axis 'micro focus' is larger than the Abby's diffraction limit. In addition, the $f_1$-cross section shows larger FWHM than $f_2$-cross section. This is because the tilted line of 'M' has a generally larger varying $a(y)$ along its structure, and the alignment of intensity through Eq. (8) is subjected to a focal length modifier $s(y)$. As a result, there exists larger focal shifts for a number of 'micro focuses' which further rendered a larger FWHM, while the intensity of whole structure is also aligned at similar level through this approach.

**Design and characterization of 'U' shaped focus.** Shown by Fig. 1 (d), a good curve-shaped focus cannot be generated simply from altering $a(y)$ through Eq. (5). Even though Eq. (8) provides enough tunabilities on focus shaping and intensity, the creation of a curve-shaped focus requires a dynamic optimization process for $a(y)$ and $s(y)$, which is complicated and time consuming.

While a simpler approach can be applied in pursuit of curve-shaped focus, here we propose a method that generate focus pattern through grating-based metalens with basic mechanisms shown in Fig. 4 (a). In Fig. 4 (a), we demonstrated an off-axis focused metalens in 2D. Since phase shifters deflect light individually, the focus remains if half of its phase shifters are discarded. A ring-shaped focus can be generated by rotating this structure (Fig. 4 a) along its central axis (curve structure with certain curvature can be generated by rotating this structure with certain angle).

Based on this mechanism we have designed an metalens with 'U'-shaped focus. This metalens is built with gratings and the basic structure fabricated is shown in Fig. 4 (b). The size of the metalens is 40×60 µm with feature size around 40 nm and thickness of 600 nm. A 'U' shaped focus is formed 15 µm away from lens plane. The characterized focal plane is shown in Fig. 4 (c), and the focus profile of two cross-sections is shown in Fig. 4 (d). A clear 'U' shaped focus can be observed on Fig. 4 (c), under 685 nm circularly polarized incidence. For this grating-based focus-pattern metalens, $a(y)$ remains the same for both vertical lines and the curve of 'U'. This is the reason that both FWHM at $f_3$ and $f_4$ cross-sections are nearly the same (Fig. 4 d). Note that the curve focus created by grating structure is polarization sensitive, and an equilibrium distribution of intensity along the curve can only be achieved under circular-polarized incidence.

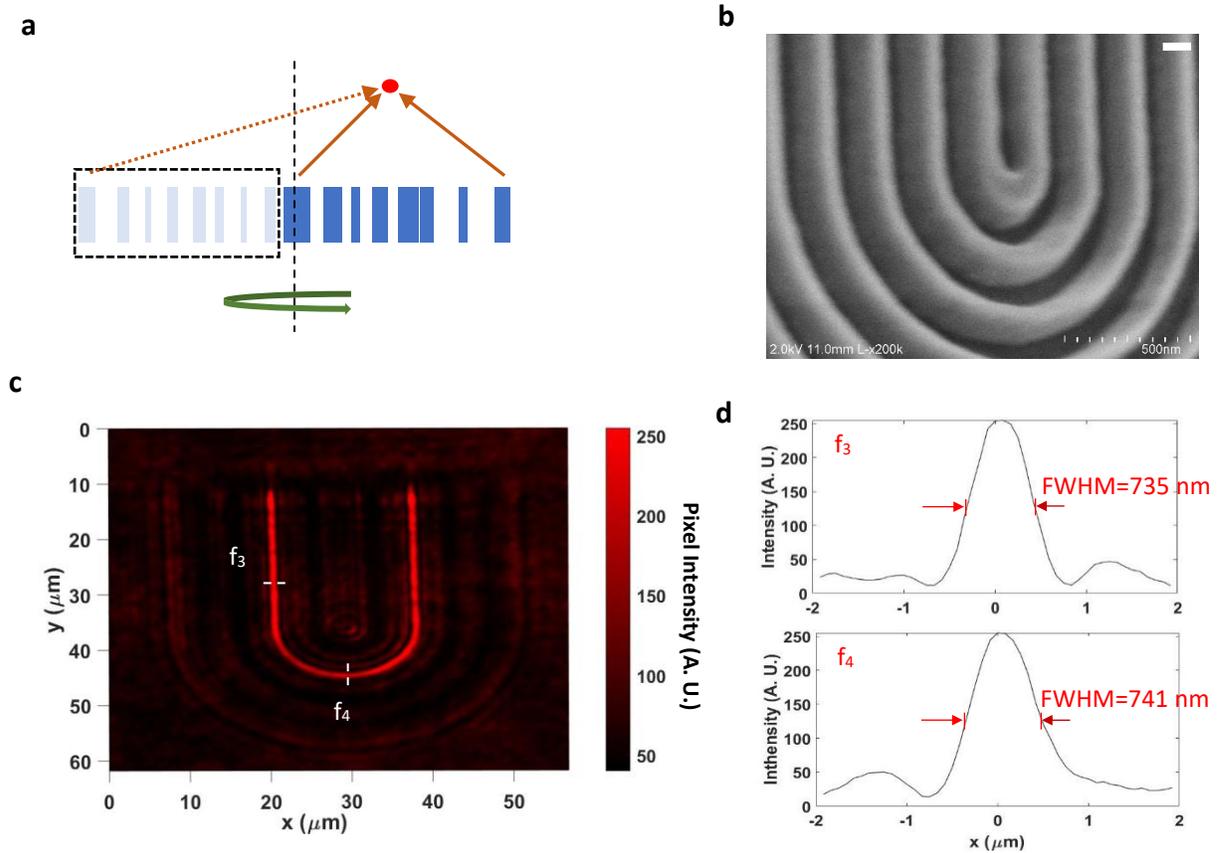

**Fig. 4** Structure and characterization of metalens with 'U' shaped focus. (**a**) Design mechanism of the curve-shaped focus. (**b**) SEM image (viewing angle tilted 30°) of the 'U' shape focused metalens, size of scale bar is 100 nm. (**c**) Characterized focal plane with incidence of 685 nm red light, two cross-sections are specified on the 'U' shaped focus. (**d**) Focus profile at two cross-sections specified respectively.

## Discussion

We have proposed and demonstrated a new approach to create metalens that focus light to an artificially designed pattern instead of a point. Two basic mechanisms of generating artificial focus patterns are proposed. Metalens with 'M' and 'U' shaped focus is fabricated and characterized for the proof of concept. During our study, we have discovered that metalens' inherent phase discontinuity has major influence on the performance of focus pattern. This results in the modification of phase construction equation with space-variant focal length modifier in pursuit of homogeneous focused intensity. New equation (Eq. 8) is proposed to achieve flexible designed focus pattern, but the idea of overlapping multiple 'micro focus' to generate focus pattern can be applied to various design as long as the requirements discussed is fulfilled.

Lens with artificial focus pattern can potentially benefit a plethora of applications that involve beam engineering. A wide range of patterns can be generated by combination of basic elements, such as 'point', 'line' and 'curve' that bring extra degrees of flexibility to beam shaping. In addition, due to the large energy density that concentrates on the pattern, metalens of this kind can potentially be applied to next generation nano-lithography with targeted pattern imprinted on the lens.

Even though the phase shifter we applied for 'M' patterned focus is cylinder-shaped, which in many cases is considered as polarization-insensitive [13], there still exist difference in incident polarization due to the anisotropic discretization of pitch (period). And the grating based artificial focus pattern is polarization dependent, due to major phase difference under different incident polarization.

Both designs we have demonstrated are based on the engineering of propagation phase through modification of effective index within each periodicity. Similar design can also be achieved through Pancharatnam-Berry phase shifters. Even though this study is based on single wavelength, strategies for achromatic metalens [11] design can also be applied to the phase shifters which may render achromatic artificial focus pattern.

## Methods

**FDTD simulation.** Three-dimensional finite difference time domain (FDTD) method (Omnisim, Photon Design Ltd.) is applied for all the simulation in this study. Perfect match layer (PML) is applied on all boundaries of simulation window to truncate simulation space. Refractive index of $SiN_x$ substrate and air are 2.74 and 1 respectively. Imaginary part of refractive index for glass and air are set zero, sufficient small grid size and sufficient long simulation times are applied for the simulation.

**Nano fabrication of metalens.** A silicon rich silicon nitride (SiNx, n = 2.74) [18] layer with thickness of 600 nm is deposited on the substrate of glass wafer as lens material. Detailed optical property of this material is shown in supplementary figure S2. Then a 230 nm thick $SiO_2$ layer is deposited on top of SiNx layer as a hard mask. Both depositions are achieved through plasma enhanced chemical vapor deposition (PECVD). A photoresist layer (ZEP520A) of 200 nm thicked is spin coated on top of $SiO_2$ layer. The 2-D lens pattern is written by E-beam and the pattern is created on the photoresist after development. The lens structure is then transferred into $SiO_2$ hard mask layer by reactive ion etching (R.I.E.) and the residual photoresist is stripped by $O_2$ plasma stripper. The pattern is finally transferred into silicon rich silicon nitride layer by another reactive ion etching process. The selectivity of $SiNx/SiO_2$ is around 2.0 with the application of 20% of $SF_6$ concentration during reactive ion etching.

The reason for our utilization of $SiO_2$ layer as hard mask is because there is no E-beam photoresist available (at 200 nm thickness) to provide enough selectivity versus SiNx to achieve direct R.I.E. with 600 nm depth. While the thickness of E-beam resist is limited by our feature size of 40 nm. As a result, a two-step R.I.E. process with hard mask is necessary for the fabrication of this metalens.

**Characterization of metalens.** A schematic diagram of characterization setup is shown in Fig. (S3). Diode laser of 685 nm light (LP685, Thorlabs Inc.) is applied as light source. The metalens sample is mounted on 3 directional motion system MT3A (Thorlabs Inc.) which is directly illuminated by the laser. The image is then magnified by an objective (Olympus LMPlanFLN 50x) and tube lens (Thorlabs TTL 200) system before captured by camera (Thorlabs 8051C, 8 megapixel).

The laser shed on the sample lens with spot size of around 10 mm which is much larger than the size of lens. When we focus on a 100x100 micrometers area, the incident source is considered as plane wave rather than a Gaussian beam. Before the fabrication of micro lens, several Au alignment marks (50x50 micron) are fabricated on the silica wafer with thickness around 20 nm. Those alignment marks will not only help us to find our structure in optical systems but also act as an important reference of scale. We first find the location of the micro lens and its nearest alignment mark. Then with fine adjustment of the sample stage (Thorlabs MT3A) in *z* direction, the objective is focused on the alignment mark. The characterization process starts with recording the *z* coordinate of the lens surface by focusing light at the alignment marks. Next step is to tune the *x-y* coordinate to move the field of view to lens area. Then the lens is moved carefully away from the objective while looking for the sharpest focus pattern generated by metalens. The focus is found where the focused pattern presenting maximum intensity. The focal length is acquired by calculating the distance moved from the lens surface to the focus in *z* direction. For the calculation of FWHM, because the pixel intensities are discreet, the exact pixel with half intensity of maximum do not exist. The FWHM presented here is calculated by taking linear approximation of adjacent pixels around half maximum intensity.

The measurement error mainly generates from: 1. The resolution of translation system (Thorlabs MT3A) is 0.5 micro meter. 2. When determine the *z* coordinate of the lens surface, the Au alignment mark has thickness of 0.02 micro meter, thus create an error of 0.02 micro meter (in worst case). 3. The sensitivity of the CMOS sensor.

## Acknowledgements

The authors would like to thank Lurie Nano Fabrication (LNF) for the distinguished nano fabrication facility provided. And thank University of Michigan and Demetra Energy for funding.

## Author contributions

Mao Ye conceived the idea. Mao Ye and Vishva Ray performed nano fabrication. Mao Ye and Dachuan Wu achieved the optical characterization. Whole process is under supervision of Ya Sha Yi.

## Additional information

**Supplementary information** accompanies this paper at the end of the manuscript.

**Competing interests:** The authors claim that they have no competing financial interest.

**Data and materials availability:** All data needed to evaluate the conclusions in the paper are present in the paper and/or the Supplementary Materials. Additional data related to this paper may be requested from the authors.

# Supplementary Materials
*Supplementary Figures*

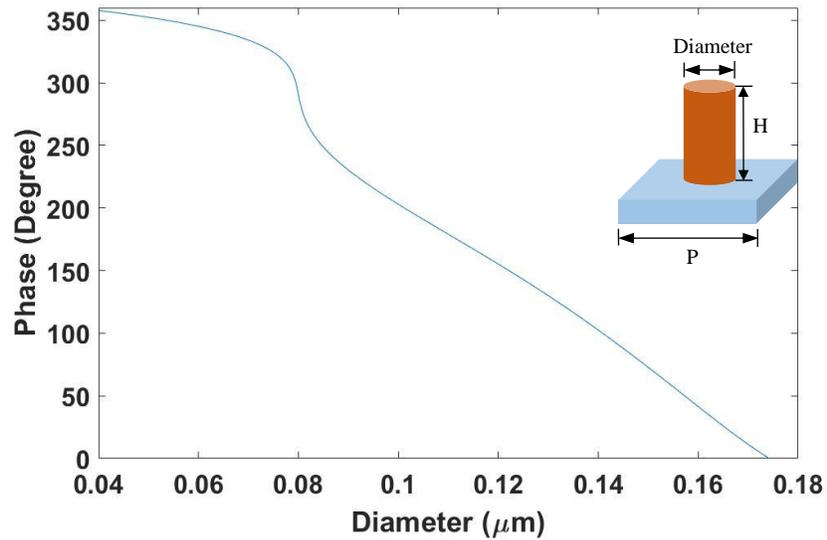

**S1.** Phase relation of nano cylinder phase shifters. The nano cylinders are designed with SiNx material (n=2.74, k=0) [18], under 220 nm of square-shaped period (P), 600 nm thickness (H).

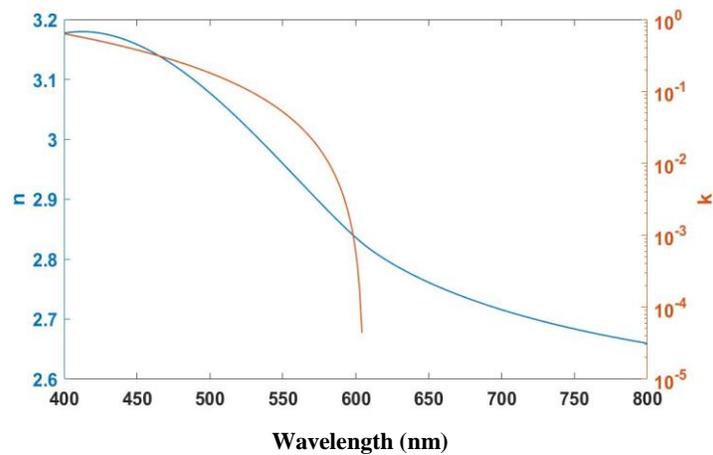

**S2.** Optical property of metalens material (silicon rich silicon nitride). n=2.74, k=0 at 685 nm. This material is deposited through PECVD. Detailed approach of getting this material can be found in reference [18].

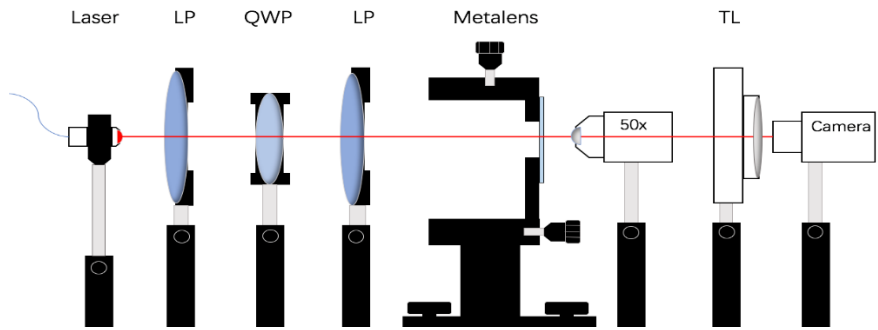

**S3.** Characterization setup. Both metalens is characterized through a laser based optical system consist of diode laser, quarter waveplate (QWP), linear polarizer (LP), 3-axis motion mount, 50x objective, tube lens (TL) and camera.